\documentclass[%
reprint,
superscriptaddress,
 amsmath,amssymb,
 aps,
prb,
]{revtex4-2}

\usepackage{graphicx}
\usepackage{dcolumn}
\usepackage{bm}
\usepackage{hyperref}
\usepackage[utf8]{inputenc}

\begin{document}

\title{Electron and Light Induced Stimulated Raman Spectroscopy for Nanoscale Molecular Mapping}

\author{Amr A. E. Saleh}
\altaffiliation{Contributed equally to this work}
\affiliation{Materials Science and Engineering Dept., Stanford University, Stanford CA 94305, USA}
\affiliation{Department of Engineering Mathematics and Physics, Faculty of Engineering, Cairo University, Giza 12613, Egypt}

\author{Daniel K. Angell}
\altaffiliation{Contributed equally to this work}
\affiliation{Materials Science and Engineering Dept., Stanford University, Stanford CA 94305, USA}

\author{Jennifer A. Dionne}
\affiliation{Materials Science and Engineering Dept., Stanford University, Stanford CA 94305, USA}
\affiliation{Department of Radiology, Molecular Imaging Program at Stanford (MIPS), Stanford University School of Medicine, Stanford, CA 94305, USA}


\begin{abstract}

We propose and theoretically analyze a new vibrational spectroscopy, termed electron- and light-induced stimulated Raman (ELISR) scattering, that combines the high spatial resolution of electron microscopy with the molecular sensitivity of surface-enhanced Raman spectroscopy. With ELISR, electron-beam excitation of plasmonic nanoparticles is utilized as a spectrally-broadband but spatially-confined Stokes beam in the presence of a diffraction-limited pump laser. To characterize this technique, we develop a numerical model and conduct full-field electromagnetic simulations to investigate two distinct nanoparticle geometries, nanorods and nanospheres, coated with a Raman-active material. Our results show the significant ($10^6$-$10^7$) stimulated Raman enhancement that is achieved with dual electron and optical excitation of these nanoparticle geometries. Importantly, the spatial resolution of this vibrational spectroscopy for electron microscopy is solely determined by the nanoparticle geometry and the plasmon mode volume. Our results highlight the promise of ELISR for simultaneous high-resolution electron microscopy with sub-diffraction-limited Raman spectroscopy, complementing advances in superresolution microscopy, correlated light and electron microscopy, and vibrational electron energy loss spectroscopy.

\end{abstract}

\maketitle

\section{Introduction}
Both electron microscopy and Raman spectroscopy have proven to be instrumental characterization tools for high spatial resolution and high molecular specificity, respectively. Modern electron microscopy provides atomic-scale spatial resolution spanning cryogenic, in-situ, and in-operando modalities~\cite{minor2019cryogenic,zheng2015frontiers}. Unprecedented structure-function relations have been revealed~\cite{nellist1995resolution,jiang2018electron,li2017atomic,baldi2014situ} yet the grey-scale images usually lack information about the sample's local chemical and molecular composition.~\cite{koster2003electron}. On the other hand, the superb specificity of Raman spectroscopy has enabled chemical characterization and identification of a variety of specimens, spanning single molecules, atomic and molecular monolayers, cells, and tissues~\cite{haran2010single,blackie2009single,kneipp1997single,wang2013raman,harz2009vibrational,pallaoro2015rapid,patel2008barcoding,kong2015raman}. The spectral fingerprint of Raman has also enabled chemical interrogation of food products, pharmaceuticals, agrochemicals, and microplastics~\cite{yang2011applications,cheng2015vibrational,langer2019present,dietze2016femtosecond,andreou2015detection,araujo2018identification}.

\begin{figure}
\begin{center}
\includegraphics{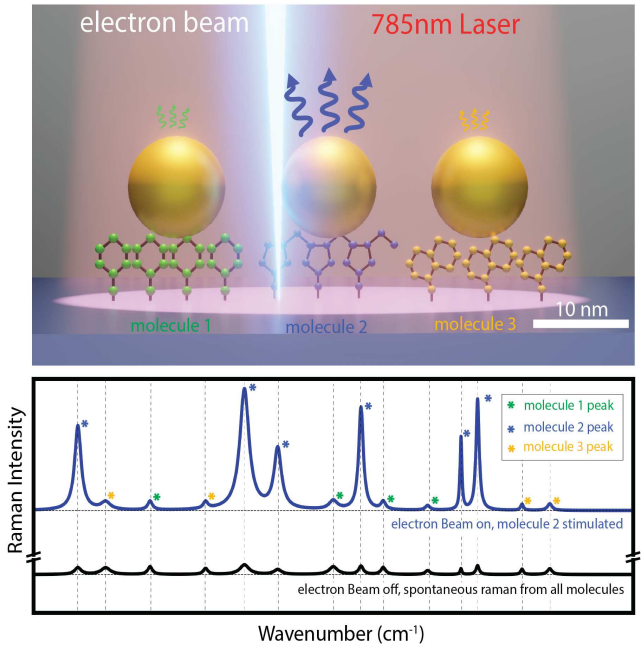}
\end{center}
\caption{(a) Schematic of ELISR. In the absence of the electron beam excitation, the Raman scattering from all illuminated molecules contributes to the spectra, prohibiting spatial localization below the diffraction limit (black illustrative spectrum). When the electron beam excites the plasmonic mode of one of the nanoparticles, the Raman scattering from the surrounding molecules is significantly enhanced while the scattering from the other molecules around unexcited nanoparticle remain unchanged (blue illustrative curve). Therefore, the molecular composition of a sample can be mapped with nanometer resolution.}
\label{fig:Figure1}
\end{figure}

To bring molecular specificity to electron microscopy, several innovative techniques have been developed, including high-resolution electron energy loss spectroscopy (EELS)~\cite{egerton2011electron,ibach2013electron,lazar2003materials} and correlated-light and electron microscopy (CLEM)~\cite{de2015correlated,hauser2017correlative}. EELS is used to probe the atomic and the chemical composition of a sample by measuring the energy distribution of the accelerated electrons upon interaction with the sample. Since the electron energy loss signal is proportional to the local dielectric constant, is can also be used probe the optical properties of materials and nanostructures~\cite{polman2019electron,yankovich2019visualizing,de2008probing,hohenester2009electron,horl2017tomographic,cao2015electron}. Most recently, it has become possible to utilize EELS to probe vibrational modes; for example, EELS has been used to probe phonon modes of thin dielectric layers~\cite{krivanek2014vibrational,konevcna2018vibrational,venkatraman2019vibrational}, surface modes of magnesium oxide nanocubes~\cite{lagos2017mapping}, as well as isotopic shifts in carbon-oxygen stretching modes~\cite{hachtel2019identification}, all with 6-10meV spectral resolution. Wide-spread adoption of this technique, however, is challenged by the required high energy resolution, which can be only achieved with advanced electron sources and monochromators, in addition to sophisticated electron detectors and spectrometers~\cite{polman2019electron}. CLEM is another approach to extract molecular information from electron microscopy images, albeit without directly probing molecular vibrations. Here, the sample is first imaged using super-resolution fluorescence microscopy to obtain molecular information, and then imaged with electron microscopy, to locate the molecular constituents within the ultra-structure of the sample~\cite{giepmans2005correlated,shu2011genetically}. Though this technique can generate remarkable overlaid datasets,  it requires nanometer-scale alignment between the electron and optical images as well as challenging and often incompatible sample preparation for both light and electron microscopy~\cite{hauser2017correlative,begemann2016correlative}.

In this paper, we introduce for the first time a novel vibrational spectral imaging technique, termed ``electron- and light-induced stimulated Raman (ELISR)'' spectroscopy. ELISR combines the molecular specificity and high-efficiency of surface-enhanced stimulated Raman scattering (SRS)~\cite{freudiger2008label} with the high spatial resolution of electron microscopy as illustrated in the schematic of Fig.~\ref{fig:Figure1}. Unlike typical stimulated Raman spectroscopy which utilizes two lasers as the pump and Stokes beams, we leverage the electron beam as a highly-localized nanoscale light source. Specifically, we utilize cathodoluminescence from plasmonic nanoparticles~\cite{vesseur2007direct,de2010optical} as the Stokes excitation in the presence of a monochromatic pump laser source. To serve as the Stokes excitation, the plasmon resonance is tuned to be red-shifted from the pump laser. When an electron beam excites a nanoparticle, only the molecules residing within the nearfield of this particular nanoparticle undergo stimulated Raman scattering while other molecules only show spontaneous Raman scattering (Fig.~\ref{fig:Figure1}). By selectively exciting individual nanoparticles using the electron beam, it becomes possible to probe the local Raman signatures of a sample with a spatial resolution determined by the mode volume of the plasmon and excitation extent of the electron beam. Consequently, ELISR has the potential to enable simultaneous electron imaging with sub-diffraction-limited Raman spectral mapping. To characterize the ELISR technique, we use full-field calculations to study metallic nanorods and nanospheres decorated with Raman-active media as model systems.

\section{ELISR modeling and numerical calculations}

In general, Raman scattering from a polarizable material, such as a molecule, can be modeled through the dynamic modulation of its polarizability~\cite{boyd2003nonlinear}. To analyze the ELISR mechanism, we investigate the interplay between this dynamic modulation and both the pump and the Stokes fields. We are specifically interested in studying how the scattered fields are altered by the presence of such polarizability modulation. Here, the molecular polarizability can be written as~\cite{shen1965theory} $\alpha = \alpha_0 + \left(\partial\alpha / \partial q \right)_0 q(t) $ where $\alpha_0$ is the equilibrium polarizability and $q(t)$ is the internuclear displacement, which can be described as a simple harmonic oscillator dynamically modulated by an external force. In the presence of an external electromagnetic field $\bm{E}$, the total force driving the vibrational mode of the molecule can be expressed as $\bm{F} = \frac{1}{2} \left(\partial\alpha / \partial q \right) \left| \bm{E} \right|^2$. For stimulated Raman scattering, the molecule experiences an optical pump field $\bm{E_p}$ with frequency $\omega_p$ and a Stokes field $\bm{E_s}$ with frequency $\omega_s$. At resonance, the difference between these two frequencies $\omega_p - \omega_s$ equals the frequency of the molecule vibrational mode $\omega_v$. Under this resonance condition, the driving force is proportional to $\bm{E_p}\bm{E_s}^*$ modulating the overall polarizability of the molecule and leading to coupling between the pump and the Stokes fields. The molecular vibration in this case is expressed as a Lorentzian function~\cite{boyd2003nonlinear} $ q(t) = B\left(\partial\alpha / \partial q \right) \bm{E_p}\bm{E_s}^* / (\omega_v^2 - \omega^2 - 2i\gamma \omega)$ with $B$ is a constant. 

In a medium with molecular density of $N$, the polarization can be expressed as 

\begin{equation}
\bm{P} = N \alpha \bm{E} = N\left[\alpha_0 + \left( \frac{\partial \alpha}{\partial q} \right) q(t)   \right] \bm{E}
\label{eq:Polarization_N}
\end{equation}

This modulation of molecular polarizability is reflected in the material's optical scattering, which can be determined from Maxwell's equations via:

\begin{equation}
\nabla^2 \bm{E} - \frac{1}{c^2} \frac{\partial^2 \bm{E}}{\partial t^2} = \frac{4\pi}{c^2} \frac{\partial^2 \bm{P}}{\partial t^2} 
\label{eq:Max_wave_eq}
\end{equation}

We note that in the absence of the molecular vibration $q(t)$, the second term of the polarization (eq.~\ref{eq:Polarization_N}) vanishes and there will be no coupling between the pump and the Stokes fields. On the other hand, when this molecular vibration is present, the coupling between the two fields is achieved. In this case, the polarizability can be alternatively expressed as~\cite{shen1965theory} $\bm{P} = N\left[\alpha_0 + \chi_R \bm{E_p}\bm{E_s}^*  \right] \bm{E}$ where $\chi_R$ is the Raman susceptibility of the material which is directly related to the molecular vibrational mode $q(t)$.

Thus, assuming an $e^{-i\omega t}$ time dependence of the field, the wave equation of the Stokes field at frequency $\omega_s$ can be written as~\cite{shen1965theory}:

\begin{equation}
\nabla^2 \bm{E_s}(\omega_s) - \frac{\omega_s^2}{c^2} \left( 1+4 \pi N \alpha_0 + 4 \pi \chi_R |\bm{E_p}|^2 \right) \bm{E_s}(\omega_s) = 0
\label{eq:Max_Raman}
\end{equation}

Eq.~\ref{eq:Max_Raman} indicates that an incident Stokes field can be amplified in the presence of a Raman active material. In this case, the Raman gain can be modeled as the change of the material permittivity under an optical pump as $\epsilon = \epsilon_b + 4 \pi \chi_{R} \left| \bm{E_p}\right|^2$. As depicted in Fig.~\ref{fig:Figure1}a, the Raman susceptibility around a vibrational mode can be described by a Lorentzian function.

In ELISR, the Stokes field $\bm{E_s}$ exhibits sub-diffraction-limited resolution by leveraging the coherent cathodoluminescence from plasmonic nanoparticles. Plasmonic cathodoluminescence occurs when swift electrons induce local charge oscillations within the nanoparticle, leading to optical radiation. ELISR uses this localized excitation to resemble a confined nanoscale Stokes excitation. Assuming negligible depletion of the optical pump, it is possible to evaluate the amplification of this electron-beam induced Stokes field in two steps following a coupled-wave approach~\cite{shen1965theory,mcanally2016coupled,lee2004theory}. First, we calculate the enhanced optical pump intensity $|\bm{E_p}|^2$ inside the Raman active material under plane wave pump excitation, and use this pump intensity to determine the change in the material permittivity. Second, we use this modified material permittivity to calculate the near-field and the far-field scattered from the nanoparticle under electron beam excitation. Using this model, we quantitatively analyze both the Raman gain (hereafter, ``gain'') and the relative enhancement of the stimulated Raman over the spontaneous Raman scattering (hereafter, ``$G_{stim}$''). Raman gain is defined as the amplification of the farfield scattering intensity at the Raman peak wavelength; it is calculated as the ratio of the total farfield scattering intensity under electron beam excitation with and without the optical pump. The enhancement of the stimulated to the spontaneous Raman scattering, $G_{stim}$, quantifies the role of the electron beam in increasing the number of the Raman photons scattered over the spontaneous Raman scattering. $G_{stim}$ can be calculated as the ratio of the differential stimulated to spontaneous Raman cross sections and is given as~\cite{mccamant2003femtosecond,wickramasinghe2014billion}: 

\begin{equation}
G_{stim} = \frac{32 \pi^2 c^2}{\omega_s^2}F(\omega_s)
\label{eq:Gstim}
\end{equation}

where $F(\omega_s)$ is the Stokes photon flux generated by the electron beam excitation of the plasmonic nanoparticle.

To compute the Raman gain and the Raman enhancement achieved with ELISR, we develop a numerical approach based on the boundary element method (BEM)~\cite{de2002retarded} implemented in Matlab (MNPBEM toolbox)~\cite{hohenester2012mnpbem,hohenester2014simulating,losquin2015unveiling}. As we indicated earlier, calculations are performed in two steps: first we calculate the electric field intensity inside the Raman-active media under plane wave excitation at the wavelength of the pump laser. We use this field intensity to modify the Raman material permittivity accordingly. Second, we use the modified permitivity to calculate the total farfield intensity under excitation with the electron beam.

\section{Results and discussion}
\subsection{Nanorods as ELISR labels in the NIR}
\begin{figure}
\begin{center}
\includegraphics{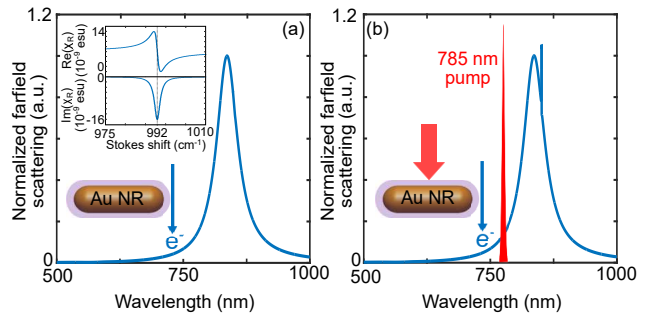}
\end{center}
\caption{(a) The normalized total farfield scattering spectrum from a 125 nm by 25 nm gold nanorod (NR) coated with 4 nm of the Raman active material under electron beam excitation in the absence of the optical pump. The inset shows the Raman susceptibility of the Raman active material used in our study with a Raman peak at 992 cm$^{-1}$. (b) In the presence of a 785 nm optical pump, stimulated Raman peak appears superimposed on the cathodoluminescence spectrum of the nanorod.}
\label{fig:Figure2}
\end{figure}

The first model system we study is a metallic gold nanorod, which has a diameter of 25nm, a length of 125nm, and is coated with a 4nm Raman-active dielectric layer. The background dielectric constant of this layer is 2.1 with a Raman response given by the Raman susceptibility ($\chi_R$) shown in the inset of Fig.~\ref{fig:Figure2}a. Such nanorod dimensions are selected so that its plasmonic resonance is red-shifted from the 785 nm Raman pump wavelength. Figure ~\ref{fig:Figure2}a shows the total farfield intensity spectrum when the nanorod is excited with a 1 nA and 80 keV electron beam positioned 2 nm away from the 4nm dielectric shell;  as seen, the nanorod has a resonant peak at 837 nm. Fig.~\ref{fig:Figure2}b depicts the total farfield scattering of the nanorod when a 10mW/$\mu$m$^2$ optical pump excitation at 785 nm is introduced along with the electron beam excitation. Here, the pump beam has introduced noticeable gain into the system at the Raman mode, which manifests itself as a Raman peak on top of the cathodoluminescence peak at 851 nm, and corresponds to a 992 cm$^{-1}$ Raman shift from the optical pump.  

\begin{figure}
\begin{center}
\includegraphics{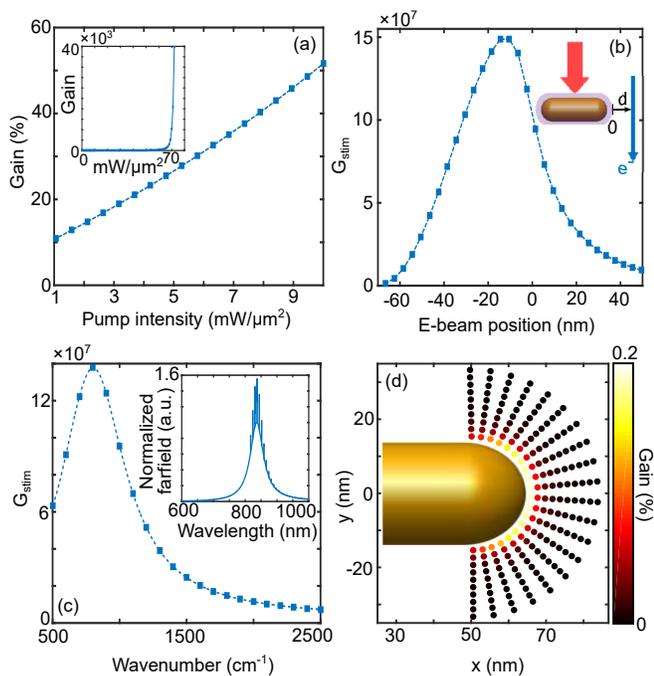}
\end{center}
\caption{(a) The Raman gain calculated as the ratio between the farfield scattering intensity at the Raman mode (851 nm) with and without the optical pump under various pump intensities. The inset shows the lasing effect when the nanorod is strongly pumped. (b) The enhancement of the stimulated to spontaneous Raman scattering as a function of the electron beam position. Here, the point of origin is located at metal tip of the coated gold nanorod. The dependence of the enhancement of the stimulated to spontaneous Raman scattering on the Raman peak wavelength (c) shows maximum enhancement achieved when the the Raman peak wavelength matches that of the nanorod plasmon resonance. The inset shows the total normalized farfield scattering spectrum with various Raman peak wavelengths. (d) The Raman peak height as a function of position around the nanorod calculated using 2 nm sphere diameter of active material located at different positions.}
\label{fig:Figure3}
\end{figure}

To evaluate the Raman gain, we calculate the ratio of the far-field scattering intensity at 992 cm$^{-1}$ with and without the optical pump, as a function of the pump power. Fig.~\ref{fig:Figure3}a shows the linear dependence of the gain on the pump power~\cite{min2011coherent}. For the model Raman material we are considering, the figure indicates that even at pump intensities as low as 1 mW/$\mu$m$^2$, the Raman gain achieved is approximately 11$\%$ and increases to more than 50$\%$ at 10 mW/$\mu$m$^2$. As shown in the inset, if the system is pumped to even higher powers, it reaches a lasing threshold around 60mW/$\mu$m$^2$.  

Besides the pump power, the electron beam position can impact generation of the Raman gain signal. As shown in cathodoluminescence and EELS studies, the efficiency of the plasmon mode excitation is specified by the impact position of the electron beam~\cite{scholl2012quantum,nelayah2007mapping}; this efficiency directly impacts the ELISR gain. In Fig.~\ref{fig:Figure3}b, we demonstrate the importance of the electron beam position on generating Raman gain. The figure plots the enhancement of the stimulated to the spontaneous raman signal ($G_{stim}$) as a function of the impact position of the electron beam; here, the beam position is measured from the tip of the coated gold nanorod and spans the longitudinal axis. As seen, $G_{stim}$ exhibits a maximum with the electron beam positioned at the metal tip of the rod, where electron-beam coupling to the nanorod plasmon is maximized. This signal decays to half its value by 5nm away from the dielectric layer, indicating that the spatial resolution of this technique is determined by the coupling efficiency between the electron beam and the nanorod at different electron beam impact positions. We also investigate the detuning of the Raman mode with that of the nanorod resonance, while keeping the pump beam at 785nm (see Fig.~\ref{fig:Figure3}c). As expected, the spectral overlap of nanorod resonance with that of the Raman mode is crucial. The graph exhibits the strongest peak (as shown in the inset of Fig.~\ref{fig:Figure3}c), and hence the largest Raman gain, when the nanorod resonance is spectrally well-aligned with the Raman vibrational mode. Therefore, while ELISR can be used for a range of molecules with Stokes shifts that overlap with the plasmon resonance, particular vibrational resonances can be selectively enhanced through spectral tuning of the nanoparticle.

The signal from ELISR is also dependent on position of the Raman active material with respect to the nanoparticle. To investigate this dependence, we replace the uniform Raman material shell with a 2-nm diameter sphere of Raman-active media and sweep this sphere over various positions from the nanorod calculating the gain achieved at these various positions. Fig.~\ref{fig:Figure3}d shows the Raman gain calculated at each position, showing that the gain is maximized when the Raman material is closest to the nanorod tip, and decays to half of its maximum value 5 nm from the tip.

This nanorod analysis highlights two important aspects of ELISR: 1), the nanoparticle geometry and associated mode volume determines the spatial resolution of this e-beam Raman probe; and 2). the nanoparticle resonance dictates the range of pump wavelengths that can be used.

\subsection{Nanospheres as ELISR labels in the visible}

\begin{figure}
\begin{center}
\includegraphics{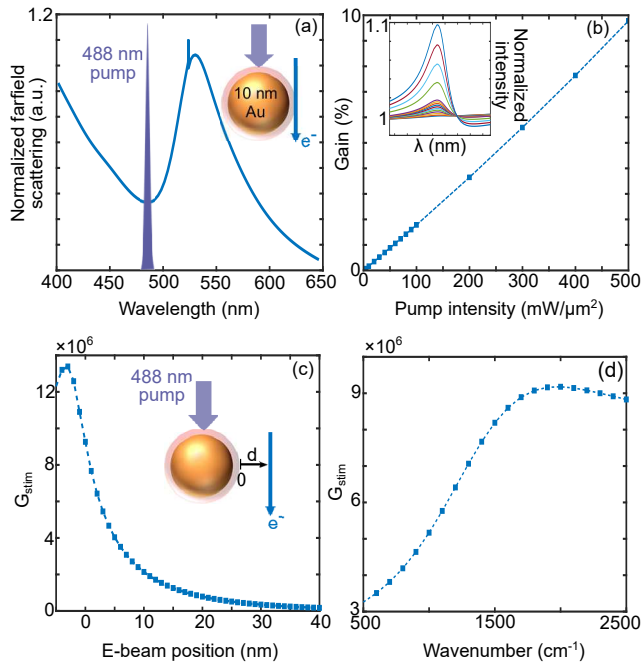}
\end{center}
\caption{(a) The normalized total farfield scattering spectrum from a 10 nm gold nanosphere coated with 2 nm of a Raman active material with a 1400 cm$^{-1}$ Raman mode under both 488 nm optical pump and  electron beam excitations. (b) The dependence of the Raman gain on the optical pump intensity. The inset shows the normalized farfield intensity at the Raman peak under the corresponding pump intensities between 523.6 nm and 523.95 nm. The enhancement of the stimulated to spontaneous Raman scattering as a function of the electron beam position (c) and the Raman peak position (d).}
\label{fig:Figure4}
\end{figure}

Smaller nanoparticle geometries, such as nanospheres, can enable higher ELISR spatial resolution. In this case, a larger number of nanoparticles can be packed on a given sample surface allowing for a larger number of ELISR measurement points per unit area. Indeed, gold nanosphere labeling is already utilized in transmission electron microscopy to improve the contrast of biological samples ~\cite{harris2006influenza}, making it attractive to investigate the performance of such geometries in ELISR. As an example for such nanospheres, We consider a 10 nm diameter gold nanosphere coated with a 2 nm Raman-active dielectric layer; the Raman susceptibility exhibits peak at 1400 cm$^{-1}$. As shown in Fig.~\ref{fig:Figure4}a, the total farfield cathodoluminescence scattering intensity spectrum shows a resonance at 530 nm when excited with an electron beam positioned 2 nm away from the nanosphere edge. When a 488 nm pump is introduced concurrent with the electron beam excitation, the Raman peak appears superimposed on the cathodoluminescence, similar to the ELISR results from the nanorod.

Fig.~\ref{fig:Figure4}b illustrates the linear dependence between the Raman gain and the pump intensity for the nanospheres, similar to that seen for the nanorods. However, the Raman gain achieved at a certain pump power is reduced compared to that of the nanorod. This reduced gain is because of the reduced extinction cross section of the nanosphere, as well as the smaller effective volume of the Raman active material interacting with the pump. In addition, because of the increased losses in gold at shorter wavelengths, the scattering efficiency from cathodoluminescence is reduced as more energy is lost in the form of heat. Consequently, the stimulated Raman enhancement is reduced by an order of magnitude compared to that of the nanorod. Fig.~\ref{fig:Figure4}c depicts $G_{stim}$ as a function of electron beam position. Similar to the nanorod, the maximum stimulated Raman enhancement is achieved when the electron beam is positioned right at the metal edge of the nanosphere. This enhancement drops to half of its maximum value when the electron beam is 5 nm away from the nanosphere. The interplay between the nanosphere resonance wavelength and the Raman peak shift wavelength is depicted in Fig.~\ref{fig:Figure4}d. As expected, and shown with the nanorod geometry, the maximum stimulated Raman enhancement is achieved when the vibrational mode frequency shift matches that of the nanosphere. Interestingly, the enhancement achieved when the vibrational mode is red-shifted from nanosphere resonance is stronger than when the vibrational mode is blue-shifted. This effect is due to the lower losses of the gold at longer wavelengths.

\section{Conclusion}
The nanorod and nanosphere highlight the benefits and constraints of ELISR:  the nanorod generates both strong nearfields and significant cathodoluminescence, yielding large Raman gain at reasonable pump powers, yet has lower spatial resolution (although still sub-wavelength). On the other hand, the nanosphere promises significantly improved spatial resolution, but requires larger pump powers in order to achieve strong stimulated Raman.

In summary, ELISR promises to enable mapping of vibrational modes at the nanoscale, concurrent with high-resolution structural imaging of the electron beam. Unlike existing methods such as EELS and CLEM, ELISR can be implemented in already available electron microscopes without the need for extreme monochromation of the electron beam, and can provide simultaneous chemical and structural information. By exploring two nanoparticles geometries, we have laid the foundation for the technique’s experimental limits and feasibility. Such a nanoscale vibrational spectroscopy technique could be accessible to any researcher with the ability to couple light into an electron microscope, including scanning and transmission electron microscopes designed for variety of cryogenic, environmental and/or in-situ and in-operando measurements. ELISR should open new opportunities to characterize the local chemical composition of a variety of samples during electron imaging, including two-dimensional materials, polymeric blends, and biological cells and tissues.

\begin{acknowledgments}
This work was supported by the U.S. Department of Energy “Photonics at Thermodynamic Limits” Energy Frontier Research Center under grant DE-SC0019140. A. S. and J. D. also acknowledge support from the Stanford Catalyst for Collaborative Solutions. D. A. was supported by the National Science Foundation Graduate Research Fellowship under Grant No. 1656518.
\end{acknowledgments}

\bibliographystyle{apsrev4-2}
\bibliography{prb_StimRaman_BibTeX_v1}

\end{document}